%% file: ms.tex
\documentclass[12pt,preprint]{emulateapj}
\usepackage{graphicx}

\input{newcommand_gen.tex} 
\input{planetinfo_156846.tex} 

\shorttitle{Transit Monitoring of HD~156846\lowercase{b}}
\shortauthors{Stephen R. Kane et al.}
\slugcomment{Submitted for publication in the Astrophysical Journal}

\begin{document}

\title{Improved Orbital Parameters and Transit Monitoring for
  HD~156846\lowercase{b}}

\author{
  Stephen R. Kane\altaffilmark{1},
  Andrew W. Howard\altaffilmark{2,3},
  Genady Pilyavsky\altaffilmark{4},
  Suvrath Mahadevan\altaffilmark{4,5},
  Gregory W. Henry\altaffilmark{6},
  Kaspar von Braun\altaffilmark{1},
  David R. Ciardi\altaffilmark{1},
  Diana Dragomir\altaffilmark{1,7},
  Debra A. Fischer\altaffilmark{8},
  Eric Jensen\altaffilmark{9},
  Gregory Laughlin\altaffilmark{10},
  Solange V. Ramirez\altaffilmark{1},
  Jason T. Wright\altaffilmark{4,5}
}
\email{skane@ipac.caltech.edu}
\altaffiltext{1}{NASA Exoplanet Science Institute, Caltech, MS 100-22,
  770 South Wilson Avenue, Pasadena, CA 91125}
\altaffiltext{2}{Department of Astronomy, University of California,
  Berkeley, CA 94720}
\altaffiltext{3}{Space Sciences Laboratory, University of California,
  Berkeley, CA 94720}
\altaffiltext{4}{Department of Astronomy and Astrophysics,
  Pennsylvania State University, 525 Davey Laboratory, University
  Park, PA 16802}
\altaffiltext{5}{Center for Exoplanets \& Habitable Worlds,
  Pennsylvania State University, 525 Davey Laboratory, University
  Park, PA 16802}
\altaffiltext{6}{Center of Excellence in Information Systems, Tennessee
  State University, 3500 John A. Merritt Blvd., Box 9501, Nashville,
  TN 37209}
\altaffiltext{7}{Department of Physics \& Astronomy, University of
  British Columbia, Vancouver, BC V6T1Z1, Canada}
\altaffiltext{8}{Department of Astronomy, Yale University, New Haven,
  CT 06511}
\altaffiltext{9}{Dept of Physics \& Astronomy, Swarthmore College,
  Swarthmore, PA 19081}
\altaffiltext{10}{UCO/Lick Observatory, University of California, Santa
  Cruz, CA 95064}


\begin{abstract}

HD~156846b is a Jovian planet in a highly eccentric orbit ($e = 0.85$)
with a period of 359.55 days. The pericenter passage at a distance of
0.16~AU is nearly aligned to our line of sight, offering an enhanced
transit probability of 5.4\% and a potentially rich probe of the
dynamics of a cool planetary atmosphere impulsively heated during
close approach to a bright star ($V = 6.5$). We present new radial
velocity (RV) and photometric measurements of this star as part of the
Transit Ephemeris Refinement and Monitoring Survey (TERMS). The RV
measurements from Keck-HIRES reduce the predicted transit time
uncertainty to 20 minutes, an order of magnitude improvement over the
ephemeris from the discovery paper. We photometrically monitored a
predicted transit window under relatively poor photometric conditions,
from which our non-detection does not rule out a transiting
geometry. We also present photometry that demonstrates stability at
the millimag level over its rotational timescale.

\end{abstract}



\section{Introduction}
\label{introduction}

The discovery of exoplanets using the transit technique is becoming
increasingly dominant amongst the various detection methods. Examples
of major contributors to the ground-based discovery of transiting
exoplanets are the Hungarian Automated Telescope Network (HATNet)
\citep{bak04} and SuperWASP \citep{pol06}. From the vantage-point of
space, the major contributors are the Kepler mission \citep{bor10} and
the CoRoT mission \citep{bar08}. The discoveries provided by these
surveys are producing insights into the exoplanet mass-radius
relationship, extending down towards super-Earth planets
\citep{sea07}. Although these space-based surveys are expected to
extend the period sensitivity to longer periods, such as the case of
CoRoT-9b \citep{dee10}, the picture is incomplete since the surveys
are strongly biased towards short-period planets around relatively
faint host stars.

Several planets discovered with the radial velocity technique have
subsequently been found to transit, the first of which was HD~209458b
\citep{cha00,hen00}. The brightness of their host stars has
facilitated further characterization of their atmospheres, such as the
cases of HD~189733b and HD~149026b (e.g., \citet{knu09a,knu09b}, see
also review article by \citet{sea10}). The Neptune-mass planet
orbiting GJ~436 became the first known transiting planet around an
M-dwarf primary \citep{gil07}. The detection of transits for the
planets HD~17156b \citep{bar07} and HD~80606b \citep{lau09,mou09},
enabled by their high eccentricities \citep{kan08,kan09a}, provided
the first insights into the structures of longer-period planets. Many
of the known planets with orbital periods larger than a few days have
yet to be photometrically monitored at predicted transit times,
hampered mostly by insufficient orbital parameter precision to
accurately predict when the planet might transit. Further discoveries
of long-period planetary transits around bright stars are vital to
understanding the dependence of planetary structure and atmospheric
dynamics on the periastron distance of the planet
\citep{for10,kan10b,lan08}. Provided the orbital parameters can be
determined with sufficient precision, monitoring planets detected via
the radial velocity technique at predicted transit times provides a
means to increase the sample of long-period transiting planets
\citep{kan09b,kan10a}. There exist efforts to detect transits of the
known radial velocity planets, such as the Spitzer search for transits
of low-mass planets \citep{gil10}. The Transit Ephemeris Refinement
and Monitoring Survey (TERMS) is a program which is capable of
monitoring long-period as well as short-period planets by refining the
orbital parameters of the system.

Here we present a detailed analysis of one such system. The massive
planet orbiting the star HD~156846 was discovered by \citet{tam08}
using the CORALIE instrument. The planet is in a highly-eccentric
orbit with a period of slightly less than a year. The periastron
argument of the orbit is such that the transit probability is
significantly enhanced compared to an equivalent circular orbit (5.4\%
compared to 0.9\%). Our combined fit to new Keck data along with the
discovery CORALIE data greatly improves the orbital parameters for the
system, allowing an accurate prediction of possible transit times. We
also find no evidence for additional companions in the system through
high-precision radial velocity (RV) data acquired during periastron
passage over multiple orbits. The long-term photometry presented here
establishes the photometric stability of the host star. We present
photometry acquired during a predicted transit window which places an
upper-limit on a transit for this planet. Finally, we discuss
additional constraints on the mass and orbit of the planet from a
potential transit null-result.


\section{Science Motivation}
\label{motivation}

Here we describe why the planet orbiting HD~156846 is a particularly
interesting target and the potential gains which may be achieved
through further studies.

HD~156846 is an extraordinarily bright star ($V=6.5$), brighter indeed
by a factor of $\sim 2.9$ than either of the planet hosting stars
HD~209458 and HD~189733. The opportunities for follow-up studies of a
fundamentally new type of planetary atmosphere would therefore be
close to optimal. Note that massive, relatively cold planets such as
this one have intrinsically difficult atmospheres to study via
transmission spectroscopy. Their atmospheric scale heights are of
order a factor of 20 smaller than typical hot Jupiters (see for
example \citet{vid11}), such that a bright host star is needed to
achieve adequate signal-to-noise.

Given the properties of this star (see Section \ref{steprop}), the
received flux of the planet at apastron will be nearly identical to
the flux received by the Earth from the Sun. It is therefore not
unreasonable to expect that the planet during this phase of the orbit
will be sheathed in reflective white water clouds. At some point prior
to periastron, when the received flux increases briefly to a value
nearly 150 times that at apastron, the received flux should be
sufficient to flash the water clouds to steam \citep{sud05}. Planets
in the post-water cloud temperature regime are expected to have
atmospheres transparent down to large pressure depths, which will
cause a dramatic drop in the planet-wide albedo. Because of the
extra-bright primary, HD 156846 will always represent one of the very
best targets in the sky for reflected light observations of this
time-sensitive albedo change, which gives real insight into the
atmospheric dynamics. This system is thus likely to gain in importance
as photometric sensitivities improve, and so any knowledge of the
inclination (and whether it transits) is very important.

\citet{ham92} have shown that the Hill radius at periastron is a good
representation of the stability zone for a satellite of a planet in
very eccentric orbit around a star. The relatively large mass of
HD~156846b leads to a sizable effective Hill Sphere at periastron
($\sim 0.02$~AU, $\sim 47$ Jupiter radii), indicating that detectable
Earth-mass satellites orbiting the planet can be dynamically stable
over the lifetime of the system if the tidal quality factor, $Q$, is
of order the Jovian value or higher \citep{bar02}. Given the mass
ratios observed for the Jovian planets and their satellites in our own
solar system, one might reasonably expect a $\sim 0.5 M_{\oplus}$
satellite, which would be readily detectable using transit timing
techniques \citep{kip09}, and perhaps even directly via space-based
photometry. Given the inflated transit probability for this planet,
and its potentially interesting dynamical history, this becomes a
prime candidate in this regard.


\section{Keck Measurements and Revised Orbital Parameters}
\label{revisedop}


\subsection{Observations}

We observed HD~156846 with the HIRES echelle spectrometer
\citep{vog94} on the 10-m Keck I telescope with the goal of improving
the accuracy of the predicted transit time to guide and temporarily
anchor a photometric monitoring campaign.  Our Keck observations
postdate the CORALIE measurements \citep{tam08} and span 2009 May to
2010 October. The \plaKNobs\ Keck RV measurements were made from
observations with an iodine cell mounted directly in front of the
spectrometer entrance slit. The dense set of molecular absorption
lines imprinted on the stellar spectra provide a robust wavelength
fiducial against which Doppler shifts are measured, as well as strong
constraints on the shape of the spectrometer instrumental profile at
the time of each observation \citep{mar92,val95}. We measured
the Doppler shift of each star-times-iodine spectrum using a modelling
procedure descended from \citet{but96} as described in \citet{how09}.
The times of observation (in barycentric Julian days), relative RVs,
and associated errors (excluding jitter) are listed in Table
\ref{tab:keck_rvs}. In cases when we observed the star 3--5 times in
quick succession, we report the mean RV and appropriately reduced
uncertainty.  We also observed HD~156846 with the iodine cell removed
to construct a stellar template spectrum for Doppler modelling and to
derive stellar properties.

\begin{deluxetable}{ccc}
  \tabletypesize{\footnotesize}
  \tablecaption{Keck Radial Velocities
  \label{tab:keck_rvs}}
  \tablewidth{0pt}
  \tablehead{
    \colhead{}         & \colhead{Radial Velocity}     & \colhead{Uncertainty}  \\
    \colhead{BJD -- 2440000}   & \colhead{(\mse)}  & \colhead{(\mse)} 
}
  \startdata
  \input{rvs_156846.tex}
  \enddata
\end{deluxetable}


\subsection{Stellar Properties}
\label{steprop}

\begin{deluxetable}{lc}
  \tabletypesize{\footnotesize}
  \tablecaption{Stellar Properties
  \label{tab:stellar_params}}
  \tablewidth{0pt}
  \tablehead{
    \colhead{Parameter}   & 
    \colhead{Value} 
  }
  \startdata
  $M_V$ & \plaMv  \\
  $B-V$  & \plaBV \\
  $V$    & \plaVmag  \\
  Distance (pc) & \plaDist \\
  $T_\mathrm{eff}$ (K) &  \plaTeff \\
  log\,$g$ & \plaLogg  \\
  \feh & \plaFeh \\
  $v$\,sin\,$i$ (km\,s$^{-1}$) &   \plaVsini \\
  $M_{\star}$ ($M_{\sun}$) &  \plaMstar \\
  $R_{\star}$ ($R_{\sun}$) &  \plaRstarIso \\
  \lrphk & \plaLogRphk  \\
  $S_\mathrm{HK}$ & \plaSval \\
  \enddata
\end{deluxetable}

We used Spectroscopy Made Easy \citep{val96} to fit high-resolution
Keck-HIRES spectra of HD~156846 (HIP~84856, TYC~6242-00339-1),
applying the wavelength intervals, line data, and methodology of
\citet{val05}. We further constrained the surface gravity using
Yonsei-Yale (Y$^2$) stellar structure models \citep{dem04} and revised
\textit{Hipparcos} parallaxes \citep{van07}, with the iterative
method of \citet{val09}. The resulting stellar parameters listed in
Table \ref{tab:stellar_params} are effective temperature, surface
gravity, iron abundance, projected rotational velocity, mass, and
radius. HD~156846 lies \plaDeltamag\ mag above the $Hipparcos$ average
main sequence ($M_V$ versus $B-V$) as defined by \citet{wri05}. These
properties are consistent with a metal-rich G0 star evolved slightly
off of the main sequence. The stellar radius, $R_{\star} =
\plaRstarIso\ R_{\sun}$, is crucial for estimating the depth and
duration of a planetary transit.

Our characterization is mostly consistent with the stellar properties
reported by \citet{tam08}. Although small, key differences are the
larger mass ($M_{\star} = 1.43 M_{\sun}$) and higher effective
temperature ($T_\mathrm{eff} = 6138 \pm 36$~K) in \citet{tam08}.  In
addition, we measured the stellar activity by measuring the strength
of the \caii\ lines which give calibrated $S_\mathrm{HK}$ values on
the Mt.\ Wilson scale and \lrphk\ values \citep{isa10}. The median of
\lrphk\ and $S_\mathrm{HK}$ values are listed in Table
\ref{tab:stellar_params} and demonstrate the HD~156846 is
chromospherically quiet, which is consistent with the photometric
stability described in Section \ref{sec:photo}.


\subsection{Keplerian Models}

\begin{figure}
      \includegraphics[width=0.47\textwidth]{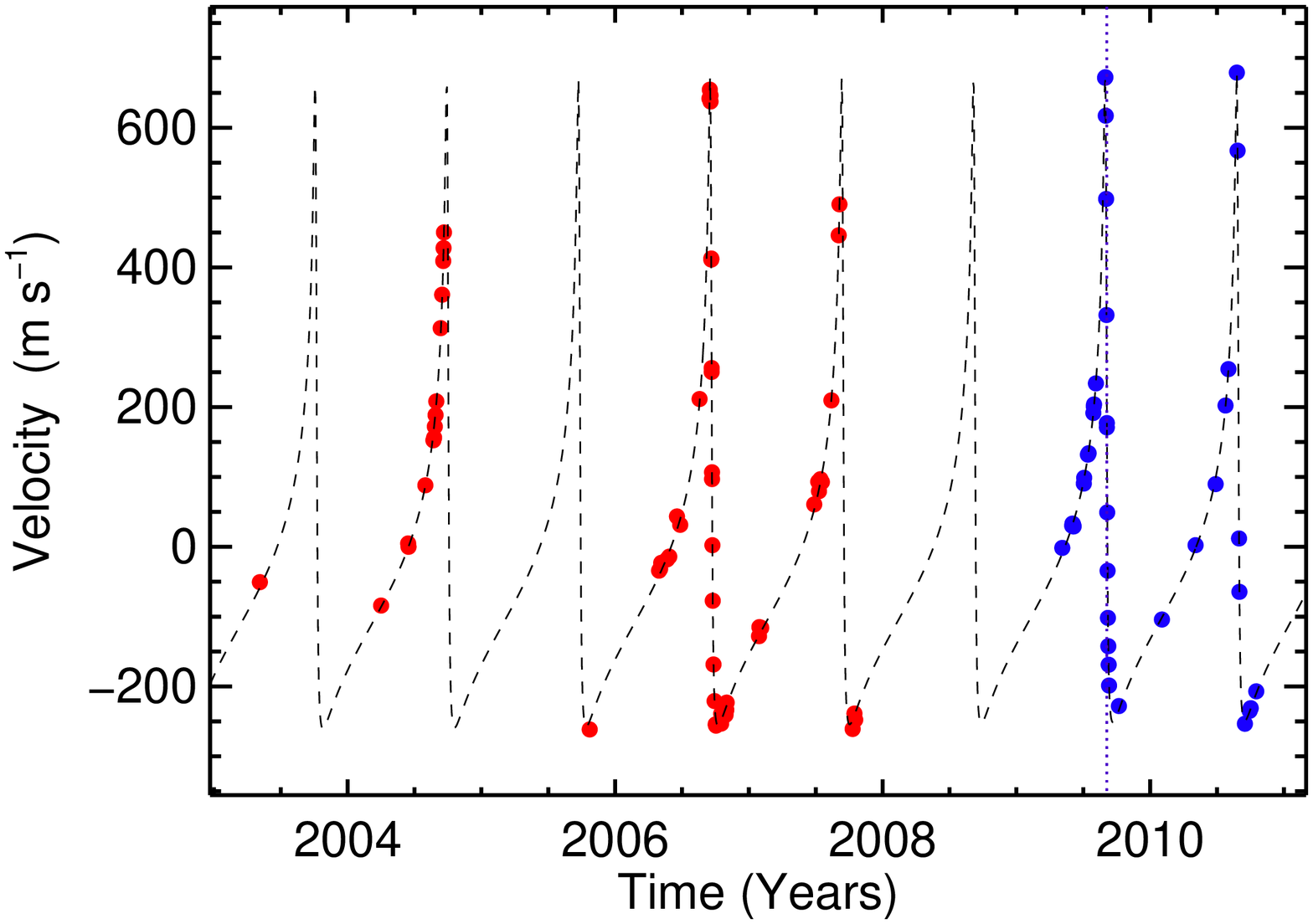} 
      \includegraphics[width=0.47\textwidth]{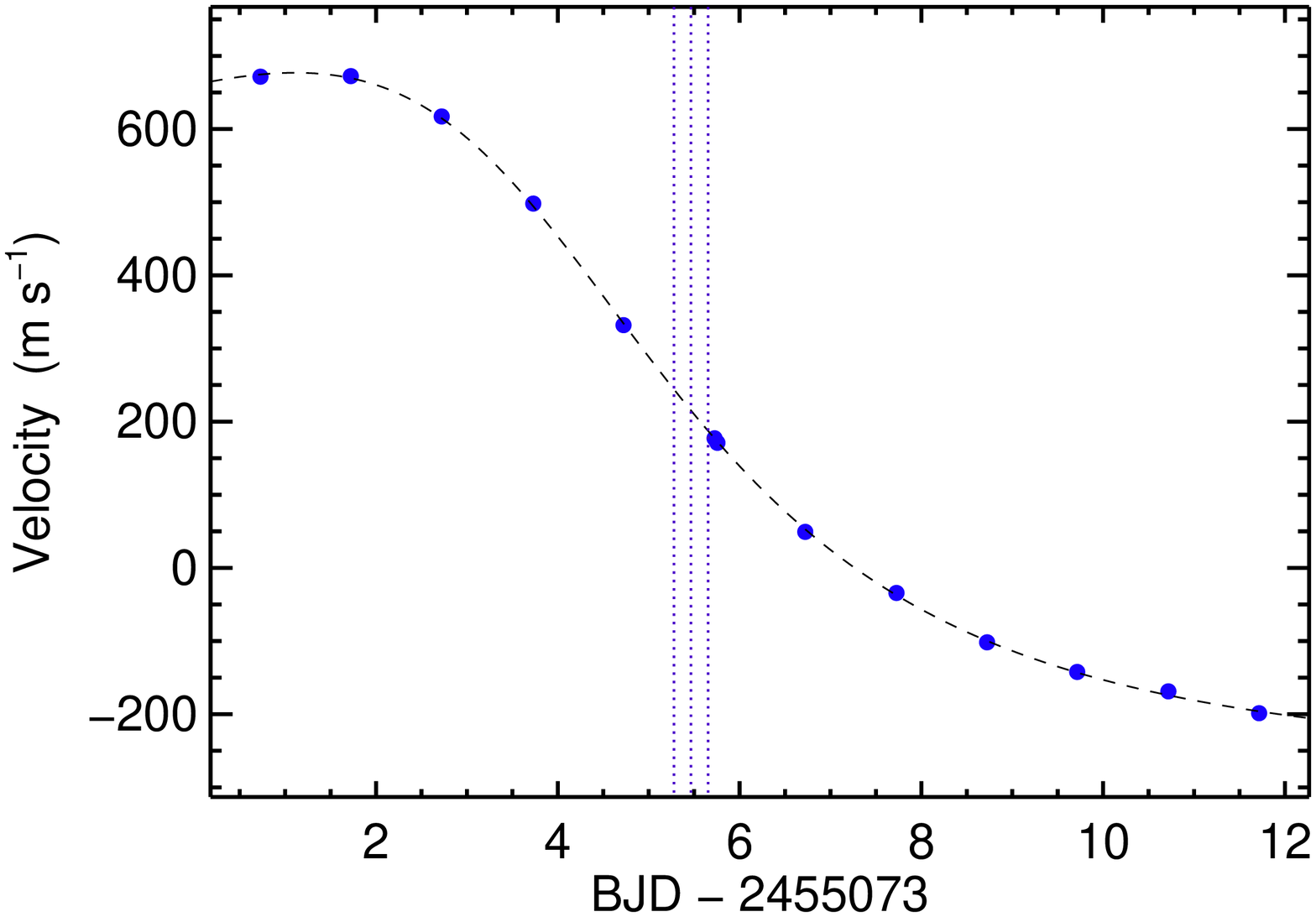} 
      \includegraphics[width=0.47\textwidth]{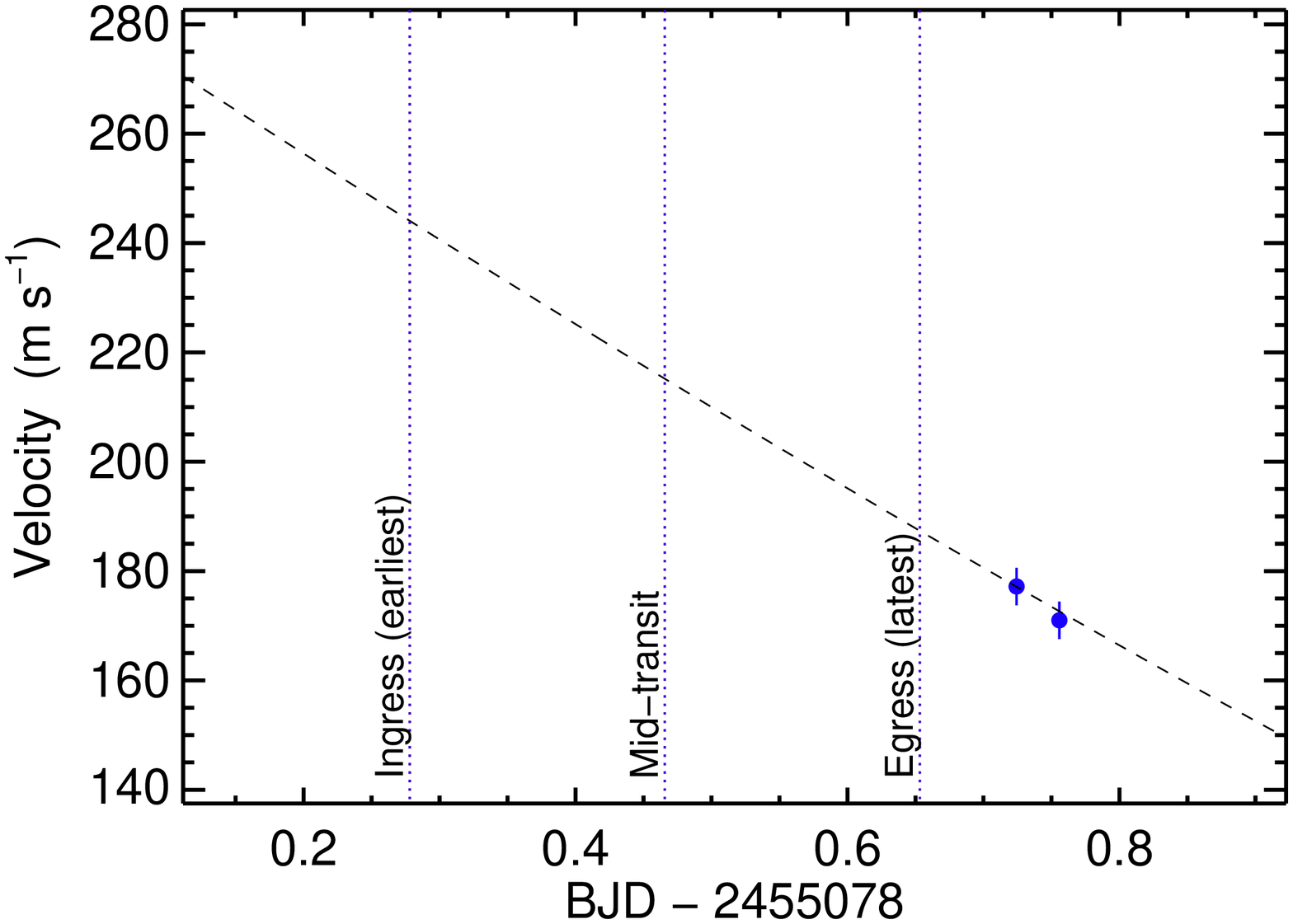} 
  \caption{Time series RV measurements of HD~156846 from CORALIE (red
    filled circles) and Keck-HIRES (blue filled circles). The adopted
    Keplerian orbital solution (Table~\ref{tab:orbital_models}) is
    shown as a dashed line in all panels. Dotted vertical lines
    illustrate the times of ingress, mid-transit, and egress for a
    predicted time of transit on 2009 September 3 (see text).  Top:
    all CORALIE and Keck-HIRES measurements.  Middle: Keck
    measurements during the 2009 September periastron passage.
    Bottom: Keck measurements on 2009 September 3.  }
  \label{fig:rvmodel}
\end{figure}

With the goal of improving the orbital solution for HD~156846b, we
constructed single-planet Keplerian models of the RVs using the orbit
fitting techniques described in \citet{how10} and the partially
linearized, least-squares fitting procedure described in
\citet{wri09}. Each velocity measurement was assigned a weight, $w$,
constructed from the quadrature sum of the measurement uncertainty
($\sigma_{\mathrm{RV}}$) and a jitter term
($\sigma_{\mathrm{jitter}}$), i.e.\ $w$ =
1/($\sigma_{\mathrm{RV}}^2+\sigma_{\mathrm{jitter}}^2$).  We chose
jitter values of $\sigma_{\mathrm{jitter}} = 3.38$ and 5.90~\ms for
Keck and CORALIE to satisfy the condition $\chi_{\nu}^2 = 1$ for
Keplerian fits to those data sets individually.  These values are
consistent with the expected jitter of a slightly evolved early G star
observed with those instruments. Sources of jitter include stellar
pulsation, magnetic activity, granulation, undetected planets, and
instrumental effects \citep{isa10,wri05}.

The Keplerian parameter uncertainties for each planet were derived
using a Monte Carlo method \citep{mar05} and account for correlations
between parameter errors. Specifically, our method accounts for
correlations between $T_c$ and the other Keplerian parameters (notably
$e$) to provide an accurate estimate of the transit time.
Uncertainties in \msini\ and $a$ reflect uncertainties in $M_{\star}$
and the orbital parameters.

\begin{deluxetable*}{lccc}
  \tablecaption{Keplerian Orbital Models
  \label{tab:orbital_models}}
  \tablewidth{0pt}
  \tablehead{
    \colhead{Parameter}   & \colhead{CORALIE}  & \colhead{Keck}  & \colhead{CORALIE$+$Keck (adopted)} 
}
  \startdata
  $P$ (days)     & \plaCPer & \plaKPer & \plaCKPer \\
  $T_c\,^{a}$ (JD -- 2,440,000) & \plaCTc  & \plaKTc  & \plaCKTc \\
  $T_p\,^{b}$ (JD -- 2,440,000) & \plaCTp & \plaKTp & \plaCKTp \\
  $e$                     & \plaCEcc & \plaKEcc & \plaCKEcc \\
  $K$ (m\,s$^{-1}$)       & \plaCK & \plaKK & \plaCKK \\
  $\omega$ (deg)          & \plaCOm & \plaKOm & \plaCKOm \\
  $dv/dt$ (m\,s$^{-1}$\,yr$^{-1}$)       & \plaCTrend & \plaKTrend & \plaCKTrend \\
  $M$\,sin\,$i$ (\mjupe) & \plaCMsini & \plaKMsini & \plaCKMsini \\
  $a$ (AU)                & \plaCA & \plaKA & \plaCKA \\
  rms (\mse) & \plaCRMS  & \plaKRMS  & \plaCKRMS \\
  \enddata
  \tablenotetext{a}{Time of transit.}
  \tablenotetext{b}{Time of periastron passage.}
\end{deluxetable*}

We considered models based on three data sets: CORALIE
\citep{tam08,tam10} and Keck-HIRES (Table~\ref{tab:keck_rvs})
individually and combined. Each model consists of a single planet in
Keplerian motion with the parameters listed in
Table~\ref{tab:orbital_models}.  We allowed for an arbitrary RV offset
in the CORALIE measurements at JD 2,454,279, the time of an instrument
upgrade. We also allowed for an RV offset between the Keck-HIRES and
CORALIE measurements.  Our models include a linear velocity trend.
Because the inclusion of a trend does not lower $\chi_{\nu}^2$, the
data do not provide evidence for a distant third body in the system.
Nevertheless, our models include the trend to provide sufficient model
flexibility to achieve properly estimated parameter uncertainties.

Table~\ref{tab:orbital_models} lists the parameters of the three, nearly
identical orbital models.  Our refitting of the \citet{tam08} data
have a slightly lower rms than they reported because we excluded
measurements with uncertainties greater then three times the median.
The higher precision Keck measurements yield a substantial improvement
in the estimated parameters. The uncertainty in the predicted time of
transit on 2009 September 3 is improved by an order of magnitude, from
0.227 days to 0.023 days. (This improvement is also due to the timing
of the Keck measurements to coincide with that epoch.) We adopt the
CORALIE$+$Keck model which has a period uncertainty of one part in
50,000 and a transit mid-point uncertainty of only 20 minutes.

Figure \ref{fig:rvmodel} shows the CORALIE and Keck measurements in
time series with our adopted model overlain. The middle panel shows
Keck measurements during the 2009 September periastron passage.  These
measurements and similar ones during the 2010 periastron passage
provide substantial leverage to constrain the orbital period and
transit time. The bottom panel shows the Keck measurements on 2009
September 3, when a transit was predicted. The dotted vertical lines
show the predicted mid-transit time and the earliest ingress and
latest egress times that are consistent with the adopted
CORALIE$+$Keck orbital solution. These times were computed from $t_c
\pm (2\sigma_{t_c} + t_{\mathrm{dur}}/2)$, where $t_c$ and
$\sigma_{t_c}$ are the predicted transit time and uncertainty, and
$t_{\mathrm{dur}}$ = 490~minutes is the duration of an equatorial
transit of the star with $R_{\star} = \plaRstarIso\ R_{\sun}$.  Our
Keck measurements that night were taken as early as possible but still
missed the transit window and do not provide a constraint on the
inclination from the Rossiter-McLaughlin effect \citep{gau07}.


\section{Transit Ephemeris Refinement}
\label{ephemeris}

\begin{figure}
  \includegraphics[angle=270,width=8.2cm]{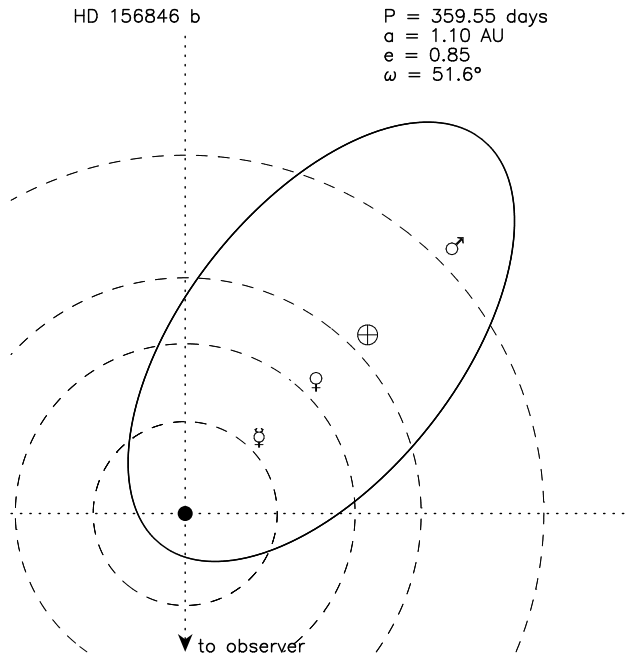}
  \caption{The orbit of the planet orbiting HD~156846 (solid line) and
    the orbits of the Solar System planets for comparison (dashed
    lines).}
  \label{fig:orbit}
\end{figure}

As described by \citet{kan08}, the transit probability of a planet is
a strong function of both the eccentricity and the argument of
periastron. In particular, the transit probability is the strongest
when the periastron passage occurs close to the star-observer line of
sight, or where $\omega = 90\degr$. The orbit of HD~156846b, shown in
Figure \ref{fig:orbit}, is well-suited for photometric follow-up at
predicted transit time since the alignment of the major axis with the
observer clearly leads to an enhanced transit probability.

From the derived stellar properties shown in Table
\ref{tab:stellar_params} and the planetary properties from Table
\ref{tab:orbital_models}, we derive a planetary radius of $R_p = 1.1
R_J$ using the methods described in \citet{bod03}. This produces a
transit probability of 5.4\% and a predicted transit depth of 3
millimags. The uncertainty in the stellar mass/radius and subsequent
uncertainty in the planetary mass/radius have a minor effect on the
estimated transit duration but in no way affects the predicted transit
mid-points since these are derived from the orbital parameters
\citep{kan09b}. Based upon the revised orbital parameters, the transit
ephemeris has been calculated for the period 2009--2016 and is shown
in Table \ref{tab:ephem}.

\begin{table*}
  \begin{center}
    \caption{Refined transit ephemeris for HD~156846b.}
    \label{tab:ephem}
    \begin{tabular}{@{}|c|c|c|c|c|c|}
      \hline
      \multicolumn{2}{|c|}{Beginning} &
      \multicolumn{2}{|c|}{Mid-point} &
      \multicolumn{2}{|c|}{End} \\
      \hline
      JD & Date & JD & Date & JD & Date \\
      \hline
      2455078.30 & 2009 09 03 19 09 & 2455078.48 & 2009 09 03 23 35 & 2455078.67 & 2009 09 04 04 01\\
      2455437.85 & 2010 08 29 08 17 & 2455438.04 & 2010 08 29 12 54 & 2455438.23 & 2010 08 29 17 30\\
      2455797.39 & 2011 08 23 21 25 & 2455797.59 & 2011 08 24 02 12 & 2455797.79 & 2011 08 24 06 59\\
      2456156.94 & 2012 08 17 10 34 & 2456157.15 & 2012 08 17 15 31 & 2456157.35 & 2012 08 17 20 28\\
      2456516.49 & 2013 08 11 23 42 & 2456516.70 & 2013 08 12 04 49 & 2456516.91 & 2013 08 12 09 57\\
      2456876.04 & 2014 08 06 12 50 & 2456876.26 & 2014 08 06 18 08 & 2456876.48 & 2014 08 06 23 26\\
      2457235.58 & 2015 08 01 01 59 & 2457235.81 & 2015 08 01 07 27 & 2457236.04 & 2015 08 01 12 55\\
      2457595.13 & 2016 07 25 15 07 & 2457595.37 & 2016 07 25 20 45 & 2457595.60 & 2016 07 26 02 23\\
      \hline
    \end{tabular}
    \tablecomments{The columns indicate the beginning, mid-point, and
      end of the transit window in both Julian and calendar date. The
      calendar date is expressed in UT and includes the year, month,
      day, hour, and minute.}
  \end{center}
\end{table*}

As described in the previous section, the size of the transit window
for 2009 was 0.37 days, or $\sim 9$ hours. The uncertainty on the
transit mid-point is small, $\sim 20$ minutes, so the transit window
is largely dominated by the transit duration. The small uncertainty on
the period ensures that the size of the transit window does not grow
substantially with time. In 2016, for example, the transit window is
only 2.5 hours longer than it was in 2009. The transit duration makes
it very difficult to attempt complete coverage of the transit window
without a multi-longitudinal campaign and exceptional weather
conditions. However, ground-based observations of either ingress or
egress will be feasible from a given observing location, provided
photometric precision requirements are met.


\section{Photometric Stability}
\label{sec:photo}

Here we describe photometry acquired outside of the transit window for
the purposes of studying the stability of the star. The only source
of time-series photometry for HD~156846 in the literature is the {\it
  Hipparcos} catalog \citep{per97}. HD~156846 (HIP~84856) was
measured 82 times over the three-year duration of the {\it Hipparcos}
mission; its variability classification in the catalog is blank,
indicating that the star ``could not be classified as variable or
constant''. The scatter (standard deviation) of the 82 observations
was 0.007 mag.

We acquired new photometry of HD~156846 with the T8 0.80~m automatic
photometric telescope (APT) at Fairborn Observatory in southern
Arizona \citep{hen99}. T8 uses a two-channel precision photometer with
two EMI 9124QB bi-alkali photomultiplier tubes to make simultaneous
measurements in the Str\"omgren $b$ and $y$ passbands. The telescope
was programmed to make differential brightness measurements of the
program star P (HD~156846, $V=6.50$, $B-V=0.58$, G1~V) with respect to
the two comparison stars C1 (HD~157379, $V=6.65$, $B-V=0.43$, F3~IV-V)
and C2 (HD~156058, $V=7.70$, $B-V=0.48$, F3~V). To improve the
precision of our differential magnitudes, we averaged the $b$ and $y$
observations to create a $(b+y)/2$ ``passband''. The typical precision
of a single observation in this combined passband is 0.0015 mag, as
measured from pairs of constant stars \citep{hen99}.

Between 2010, June 01 and July 06, the APT collected 32 good
measurements of the P$-$C1, P$-$C2, and C2$-$C1 differential
magnitudes with standard deviations of 0.00219, 0.00229, and 0.00202
mag, respectively, slightly higher than the typical 0.0015 mag
precision. However, HD~156846 and its comparison stars are located
between $-16\arcdeg$ and $-19\arcdeg$ declination, so they are
observed at higher than average air mass.  The observed scatter of the
P$-$C1, P$-$C2, and C2$-$C1 differential magnitudes are all consistent
with constant stars. Periodogram analyses of the three data sets
reveal no significant periodicity. We conclude that HD~156846 is
constant on its rotation timescale.


\section{Monitoring the Transit Window}

HD~156846 was observed at the Cerro Tololo Inter-American Observatory
(CTIO) 1.0m telescope using the Y4KCam Detector, which is a
4k$\times$4k CCD with a field of view of about 20 arcminutes on the
side\footnote{http://www.astronomy.ohio-state.edu/Y4KCam/}. We
monitored HD~156846
in the Johnson $B$ band on the nights of 3
Sep 2009 -- during its transit -- and 4, 8, 9, 11, and 12 Sep for
out-of-transit calibration purposes.
Due to
HD~156846's brightness ($V = \plaVmag; B = \plaBmag$), we
used a diaphragm in the shape of a ring constructed of plywood to
decrease the effective aperture of the telescope, thereby reducing
flux by around 40\%, similar to the technique employed by
\citet{lop06}.
For the observing strategy and photometry, our methods are similar to
those described by \citet{sou09}. To
ensure that no part of the stellar point spread function (PSF) reaches
the non-linearity levels of the CCD, and to maximize the number of
counts contained in the PSF, we defocused the telescope such that the
full-width-half-maximum of the PSF subtended 3--8 arcseconds, which
produced $\sim 10^6$ ADUs (and thus 1 millimag photon noise) per PSF
per measurement.
Observing conditions on the night of the transit window (2009
September 3) were not optimal and plagued with thin cirrus.  We
selected the brightest four reference stars in the frame and relative
photometry was performed using the methods described in \citet{eve01}.

The new calculated stellar radius presented in this paper is $2.12
R_{\sun}$, significantly larger than predictions based purely upon the
spectral type and luminosity class of the star. This has three primary
effects on the transit prediction. The first is to increase the
transit probability to 5.4\%. The second is to decrease the predicted
transit depth to 3 millimags. The third is to increase the predicted
transit duration to $\sim 9$~hours. The last two aspects are
particularly harmful to attempts at detecting a transit since they
increase the photometric precision requirements and decrease the
chances that the window can be monitored while the target is
observable.

\begin{figure}
  \includegraphics[angle=270,width=8.2cm]{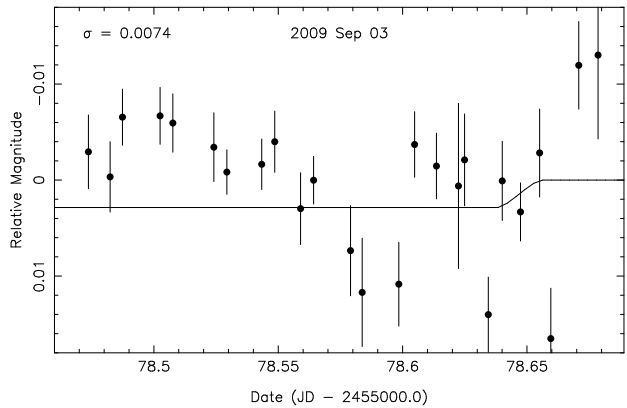}
  \caption{Binned photometry of HD~156846 from the night of the
    transit window. The overlaid solid line shows the predicted
    transit signature.}
  \label{fig:phot}
\end{figure}

Figure \ref{fig:phot} shows the photometry from the night of the
transit window. To improve the rms scatter, we binned the data into 30
equal time intervals. This improved the 1$\sigma$ scatter from
9~millimags to 7~millimags. Unfortunately, the poor conditions on that
night prevented the necessary precision from being achieved since this
is still a factor of two greater than the predicted transit depth. We
calculated the predicted transit signature based upon the analytic
models of \citet{man02}, overplotted as a solid line in the
figure. For a more detailed study on the effect of eccentric orbits on
transit lightcurves, we refer the reader to \citet{kip08} and
\citet{kip10}. Although we see no evidence for a transit in our data,
the photometric precision is inadequate to rule out such an event.
For the stellar radius we adopt, the predicted transit depth is likely
to be quite robust against variations in the planetary radius.
\citet{for07} showed that, for a given planetary composition,
planetary radii should not vary substantially between orbital radii of
0.1--2.0~AU. In order to produce a transit depth comparable to the
level of precision on the night of the transit window, the radius of
the companion would need to be $> 1.8$ Jupiter radii. Furthermore,
the increase in the estimated stellar radius caused the predicted end
of the transit to occur when the lightcurve is heavily influenced by
both high airmass and cloud contamination. Thus improved precision
would not have allowed sufficient in and out of transit data to be
acquired in order to comfortably secure the detection.

\begin{figure}
  \includegraphics[angle=270,width=8.2cm]{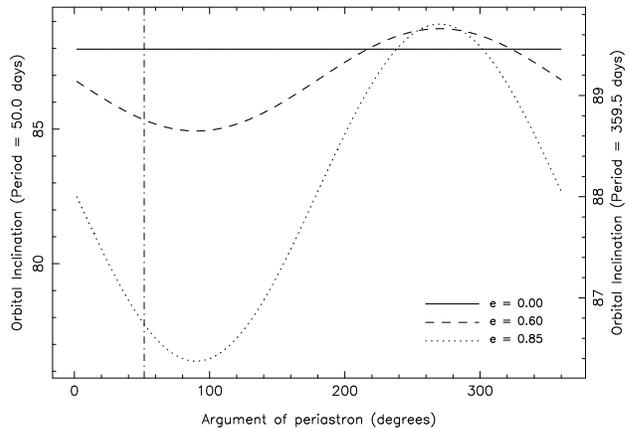}
  \caption{The maximum orbital inclination for a non-transiting planet
    as a function of the argument of periastron for eccentricities of
    0.0 (solid line), 0.6 (dashed line), and 0.85 (dotted line),
    plotted for periods of both 50.0 days and 359.5 days. The vertical
    line indicates the location of the measured periastron argument
    reported here.}
  \label{fig:inc}
\end{figure}

If a transit of this planet were to be ruled out, then weak
constraints on the inclination of the orbit could be placed. The
magnitude of these constraints as a function of periastron argument
are shown in Figure \ref{fig:inc} which shows the maximum orbital
inclination for two different periods and three different
eccentricities, including the period and eccentricity of HD~156846b. A
successful null-detection would limit the inclination to $i <
86.5\degr$ which would thus place a lower limit on the planetary mass
of 10.59 Jupiter masses. The period of 50 days is shown for
comparitive purposes, where one can rapidly improve the lower mass
limits for the smaller star--planet separation.

\section{Conclusions}

This study has been carried out as part of the Transit Ephemeris
Refinement and Monitoring Survey (TERMS). The purpose of the presented
research was to improve the orbital parameters of the known exoplanet
HD~156846b and to monitor the transit window. We present new Keck data
which, combined with previously acquired CORALIE data, refines the
orbital parameters of the planet. The measurements obtained during
periastron passage have allowed us to construct an exceptionally
accurate transit ephemeris which we present here up until the year
2016. The value in a successful transit detection would be high for
such a long-period planet in an eccentric orbit since it would provide
insight into the mass-radius relationship for planets in this regime
as well as allow follow-up characterization to determine the radiative
time scale and other properties of the atmosphere.

The challenge of attempting to monitor the transit window is
substantial since predicted transit windows are so infrequent, the
predicted transit duration is relatively long, and photometry of
bright stars becomes complicated when comparison stars are few.  We
present one such attempt here where observations were undertaken
during non-photometric conditions which does not allow the transit to
be decisively ruled out. A more suitable facility to use for such a
search is the Microvariability and Oscillations of STars (MOST)
satellite, such as that carried out by \citep{cro07}. From the ground,
the planned telescopes and instruments of the Las Cumbres Observatory
Global Telescope (LCOGT) Network \citep{shp10} will be ideal for
transit monitoring due to both their aperture size and longitude
coverage. The recent astrometry work of \citet{ref11} appears to
indicate that HD~156846b may not be in an edge-on orbit, but the
results are uncertain enough to make this study a worthwhile
exercise. We thus encourage future observations of transit windows for
this planet for those cases where the window is aligned with the
visibility of the target. Since the period of the planet is slightly
less than one year, this situation will gradually improve with each
successive transit window.


\section*{Acknowledgements}

The authors would like to thank the anonymous referee, whose comments
greatly improved the quality of the paper. This work made use of the
SIMBAD database (operated at CDS, Strasbourg, France), NASA's
Astrophysics Data System Bibliographic Services, and the NASA Star and
Exoplanet Database (NStED). This work was partially supported by
funding from the Center for Exoplanets and Habitable Worlds. The
Center for Exoplanets and Habitable Worlds is supported by the
Pennsylvania State University, the Eberly College of Science, and the
Pennsylvania Space Grant Consortium. Finally, the authors wish to
extend special thanks to those of Hawai`ian ancestry on whose sacred
mountain of Mauna Kea we are privileged to be guests. Without their
generous hospitality, the Keck observations presented herein would not
have been possible.


\end{document}

%% file: newcommand_gen.tex
\newcommand{\ms}{\mbox{m\,s$^{-1}~$}}

\newcommand{\mse}{\mbox{m\,s$^{-1}$}}

\newcommand{\mjupe}{$M_{\rm Jup}$}

\newcommand{\feh}{\ensuremath{[\mbox{Fe}/\mbox{H}]}}
\newcommand{\rphk}{\ensuremath{R'_{\mbox{\scriptsize HK}}}}
\newcommand{\lrphk}{\ensuremath{\log{\rphk}}}
\newcommand{\caii}{\ion{Ca}{2} H \& K}

\newcommand{\msini}{\ensuremath{M \sin i}}

%% file: planetinfo_156846.tex
\newcommand{\plaKPer}{\ensuremath{359.5517 \pm 0.0270}}             
\newcommand{\plaKTc}{\ensuremath{15078.470 \pm 0.017}}      
\newcommand{\plaKTp}{\ensuremath{15076.671 \pm 0.023}}      
\newcommand{\plaKEcc}{\ensuremath{0.84828 \pm 0.00061}}      
\newcommand{\plaKOm}{\ensuremath{51.42 \pm 0.18}}      
\newcommand{\plaKK}{\ensuremath{463.64 \pm 1.20}}      
\newcommand{\plaKMsini}{\ensuremath{10.54 \pm 0.29}}      
\newcommand{\plaKA}{\ensuremath{1.096 \pm 0.021}}      
\newcommand{\plaKTrend}{\ensuremath{0.86 \pm 1.36}}      
\newcommand{\plaKNobs}{\ensuremath{41}}      
\newcommand{\plaKRMS}{\ensuremath{3.45}}      
\newcommand{\plaCPer}{\ensuremath{359.5400 \pm 0.1072}}             
\newcommand{\plaCTc}{\ensuremath{15078.459 \pm 0.219}}      
\newcommand{\plaCTp}{\ensuremath{15076.699 \pm 0.227}}      
\newcommand{\plaCEcc}{\ensuremath{0.84696 \pm 0.00109}}      
\newcommand{\plaCOm}{\ensuremath{52.51 \pm 0.39}}      
\newcommand{\plaCK}{\ensuremath{464.03 \pm 2.14}}      
\newcommand{\plaCMsini}{\ensuremath{10.60 \pm 0.29}}      
\newcommand{\plaCA}{\ensuremath{1.096 \pm 0.021}}      
\newcommand{\plaCTrend}{\ensuremath{3.51 \pm 1.41}}      
\newcommand{\plaCRMS}{\ensuremath{6.85}}      
\newcommand{\plaCKPer}{\ensuremath{359.5546 \pm 0.0071}}             
\newcommand{\plaCKTc}{\ensuremath{15078.483 \pm 0.015}}      
\newcommand{\plaCKTp}{\ensuremath{15076.686 \pm 0.021}}      
\newcommand{\plaCKEcc}{\ensuremath{0.84785 \pm 0.00050}}      
\newcommand{\plaCKOm}{\ensuremath{51.62 \pm 0.16}}      
\newcommand{\plaCKK}{\ensuremath{464.14 \pm 0.96}}      
\newcommand{\plaCKMsini}{\ensuremath{10.57 \pm 0.29}}      
\newcommand{\plaCKA}{\ensuremath{1.096\pm 0.021}}      
\newcommand{\plaCKTrend}{\ensuremath{1.55 \pm 0.88}}      
\newcommand{\plaCKRMS}{\ensuremath{6.06}}      
\newcommand{\plaMv}{\ensuremath{3.055}}      
\newcommand{\plaBV}{\ensuremath{0.557}}      
\newcommand{\plaBmag}{\ensuremath{7.063}}      
\newcommand{\plaVmag}{\ensuremath{6.506}}      
\newcommand{\plaDeltamag}{\ensuremath{1.13}}    
\newcommand{\plaDist}{\ensuremath{49.0  \pm  2.2}}      
\newcommand{\plaFeh}{\ensuremath{+0.17 \pm 0.04}}      
\newcommand{\plaTeff}{\ensuremath{5969 \pm  44}}      
\newcommand{\plaVsini}{\ensuremath{5.05 \pm 0.50}}      
\newcommand{\plaLogg}{\ensuremath{3.92 \pm 0.08}}      
\newcommand{\plaMstar}{\ensuremath{1.35 \pm 0.045}}      
\newcommand{\plaRstarIso}{\ensuremath{2.12 \pm 0.12}}      
\newcommand{\plaLogRphk}{$-5.071$}      
\newcommand{\plaSval}{0.144}      

%% file: rvs_156846.tex
 14957.02010 & -111.41 &    1.94  \\
 14983.95333 &  -79.61 &    1.73  \\
 14984.87930 &  -76.60 &    1.74  \\
 14985.97062 &  -77.24 &    1.63  \\
 14986.96751 &  -78.81 &    1.69  \\
 14987.91548 &  -80.71 &    1.54  \\
 15015.85870 &  -19.24 &    1.60  \\
 15016.78653 &  -11.02 &    1.61  \\
 15026.95355 &   22.14 &    1.29  \\
 15028.96769 &   24.15 &    1.27  \\
 15041.87653 &   81.78 &    1.64  \\
 15042.95724 &   91.65 &    1.63  \\
 15043.79662 &   94.00 &    1.56  \\
 15048.77399 &  124.15 &    1.90  \\
 15073.72838 &  561.70 &    1.38  \\
 15074.72238 &  562.49 &    1.22  \\
 15075.72238 &  507.32 &    0.72  \\
 15076.73007 &  388.21 &    0.78  \\
 15077.72296 &  222.03 &    0.73  \\
 15078.72435 &   67.34 &    0.59  \\
 15078.75583 &   61.16 &    0.57  \\
 15079.72234 &  -60.59 &    0.74  \\
 15080.72602 & -144.09 &    0.49  \\
 15081.72293 & -211.59 &    1.22  \\
 15082.71457 & -252.11 &    0.86  \\
 15083.71752 & -278.63 &    0.67  \\
 15084.71590 & -308.49 &    1.13  \\
 15111.70803 & -337.84 &    1.80  \\
 15229.17709 & -213.95 &    1.67  \\
 15320.98063 & -107.50 &    1.67  \\
 15375.84930 &  -20.06 &    1.82  \\
 15402.81296 &   92.47 &    1.65  \\
 15410.78620 &  144.58 &    1.51  \\
 15433.71936 &  569.13 &    1.20  \\
 15435.71871 &  457.38 &    0.69  \\
 15439.71931 &  -97.88 &    0.75  \\
 15440.75164 & -174.32 &    0.70  \\
 15455.70798 & -363.19 &    0.72  \\
 15468.74647 & -344.65 &    1.29  \\
 15471.70625 & -340.87 &    0.77  \\
 15486.68538 & -316.77 &    0.92  \\